\documentclass[12pt]{JHEP3}
\let\ifpdf\relax
\usepackage{amsmath, amscd, amsmath, amsthm, amssymb, xy, xypic}
\usepackage{amssymb}
\usepackage{epsfig}
\usepackage{longtable}
\usepackage{url}

\numberwithin{equation}{section}

\def \hd #1 {\bfseries #1  \mdseries}
\def \italic #1 {\bfseries \it #1 \rm \mdseries}

\def \cen #1 { \begin{center} #1 \end{center}}

\def \setminus {-}

\ifpdf
\else
\usepackage{colordvi}
\fi

\numberwithin{equation}{section}

\title{Updates on Calabi--Yau manifolds from pairs of non-compact Calabi--Yau manifolds}

\author{
Nam-Hoon Lee
\\
Department of Mathematics Education, Hongik University,
42-1, Sangsu-Dong, Mapo-Gu,
Seoul 121-791, South Korea
\\
E-mail: \email{nhlee@hongik.ac.kr}
\\
}

\abstract{
We discuss developments in the construction of Calabi--Yau manifolds by smoothing normal crossing unions of quasi-Fano manifolds, following the introduction of this method to the physics community in 2010. We describe constructions of Calabi--Yau threefolds with unbounded
second Betti numbers and with very small Hodge numbers, as well as
the first non-K\"ahler examples in dimensions greater than three.
The role of Landau--Ginzburg models  in  mirror constructions is also explained.
}


\preprint{}

\begin{document}

\section{Introduction}
By a Calabi--Yau manifold, we mean a compact complex manifold $M$ with trivial canonical class and $h^{i,0}(M)=0$ for $0<i<\dim M$.
Calabi--Yau manifolds play a central role in string theory, particularly
as internal spaces for compactification and as geometric settings for
mirror symmetry. A large part of the physics literature has concentrated
on Calabi--Yau manifolds constructed as complete intersections in
projective or toric varieties, hypersurfaces in toric varieties, and
quotients of products involving elliptic curves, $K3$ surfaces, or
abelian varieties. 

There are, however, other methods of constructing Calabi--Yau manifolds
which arise naturally from degeneration theory. In \cite{Leejhep}, the
author introduced to the physics community a construction in which a
compact Calabi--Yau manifold is obtained by gluing two non-compact
Calabi--Yau manifolds. In algebro-geometric terms, the construction
begins with a normal crossing variety
$$
X_0 = Y_1 \cup Y_2,
$$
where $Y_1$ and $Y_2$ are smooth projective varieties with $h^{i,0}(Y_j)=0$ for $0<i \le \dim Y_j$, $j=1,2$, meeting along a
common smooth divisor $D = Y_1 \cap Y_2$. The divisor $D$ is assumed to be an
anticanonical divisor in both components. Thus
$$
D \in |-K_{Y_1}|
\qquad\hbox{and}\qquad
D \in |-K_{Y_2}|.
$$
If the $d$-semistability condition,
$$ N_{D/Y_1} \otimes  N_{D/Y_2} \simeq {\mathcal O}_D,$$
is satisfied, the normal crossing variety $X_0$ admits a smoothing
$$
\pi : {\cal X} \longrightarrow \Delta
$$
such that
$$
\pi^{-1}(0) = X_0
$$
and the general fiber
$$
X_t = \pi^{-1}(t), \qquad t \ne 0,
$$
is a compact Calabi--Yau manifold (\cite{KaNa}).

This construction has two complementary interpretations. From the
algebro-geometric point of view, it is a smoothing of a normal crossing
variety. This formulation makes it possible to apply logarithmic
deformation theory and to give precise conditions for the existence of
the smoothing. It also allows one to calculate invariants of the smooth
fiber from the geometry of $Y_1$, $Y_2$, and $D$.

From the differential-topological point of view, removing the common
anticanonical divisor from the two components gives
$$
Y_1^* = Y_1 \setminus D
\qquad\hbox{and}\qquad
Y_2^* = Y_2 \setminus D.
$$
These are non-compact Calabi--Yau manifolds. The compact Calabi--Yau manifold $X_t$ may then be regarded
topologically as a manifold obtained by gluing $Y_1^*$ and $Y_2^*$ near
their ends. The two descriptions therefore express the same basic idea:
a compact Calabi--Yau manifold is constructed from a pair of
non-compact Calabi--Yau building blocks. See \cite{Leejhep} for more details.

More than sixteen years have passed since this construction was
presented to the physics community. During this period, the method and
the ideas related to it have developed in several directions. Many new
families of Calabi--Yau manifolds have been constructed by smoothing
normal crossing varieties. Some of these examples have geometric or
topological properties  not  found among the standard
complete-intersection constructions.

A particularly notable development is the construction of non-K\"ahler Calabi--Yau manifolds. For instance, simply connected non-K\"ahler Calabi--Yau threefolds with unbounded second Betti numbers have been constructed. The construction
demonstrates that the topology of non-K\"ahler Calabi--Yau threefolds
can be substantially less restricted than that of their K\"ahler
counterparts. The method has also been extended to higher dimensions,
including the construction of a non-K\"ahler Calabi--Yau fourfold.

Another important development concerns mirror symmetry. A degeneration
of a Calabi--Yau manifold into two quasi-Fano components meeting along
a common anticanonical divisor is closely related to the notion of a
Tyurin degeneration (\cite{Ty}). Mirror symmetry for such degenerations suggests
that the mirror manifold should carry a fibration obtained by combining
Landau--Ginzburg models associated with the two quasi-Fano components.
This viewpoint was suggested by Doran, Harder, and Thompson and provides a geometric relation between Tyurin degenerations and fibrations on mirror Calabi--Yau manifolds \cite{DoHaTh}. These ideas have been developed to produce many examples of mirror pairs of Calabi--Yau threefolds by the smoothing method in \cite{Lee20}.

The article is organized as follows. Section 2 reviews recent constructions of Calabi--Yau threefolds obtained by smoothing normal crossing varieties. We begin with simply connected non-K\"ahler Calabi--Yau threefolds having arbitrarily large second Betti number, and then turn to examples with very small Hodge numbers, including Calabi--Yau threefolds with $h^{1,1}=h^{1,2}=1$. The higher-dimensional extensions of the smoothing method are discussed in Section 3, where we describe the first example of a non-K\"ahler Calabi--Yau fourfold and, more generally, non-K\"ahler Calabi--Yau manifolds of every dimension greater than three with unbounded second Betti numbers. Section 4 is devoted to the construction of mirror pairs of Calabi--Yau threefolds from mirror pairs of quasi-Fano threefolds, with particular attention to examples arising from three-dimensional reflexive polytopes. The final section discusses  possible future directions.

\section{Calabi--Yau threefolds with unbounded $b_2$
 and with $h^{1,1}=h^{1,2}=1$}

Friedman constructed infinitely many topological types of Calabi--Yau threefolds with $b_2=0$ and unbounded third Betti number $b_3$ (\cite{Fr}). In \cite{HaSa}, Hashimoto and Sano subsequently constructed simply connected non-K\"ahler Calabi--Yau threefolds with arbitrarily large second Betti numbers by smoothing normal crossing varieties.

The construction starts with
$$
P:=\mathbb P^1\times\mathbb P^1\times\mathbb P^1
$$
and a very general smooth K3 surface
$$
S\in |\mathcal O_P(2,2,2)|.
$$
 Projecting $S$ onto any
pair of the three $\mathbb P^1$ factors gives a double cover of
$\mathbb P^1\times\mathbb P^1$. The deck transformations of these
three double covers define three natural involutions of $S$.
These involutions generate an infinite subgroup of
$\operatorname{Aut}(S)$. The authors exploit these automorphisms to
construct infinitely many Calabi--Yau threefolds. Let $\iota$ be the
composition of two of the three involutions. Then $\iota$ is an
automorphism of infinite order.

Fix a positive integer $a$. Take two copies of $P$, and denote one of
them by $X_2=P$. The other copy is modified as follows. First, blow it
up along $a$ disjoint smooth elliptic fibers contained in $S$. 
It is then blown up along an additional smooth curve
$C_a\subset S$, chosen so that the normal bundles of the two copies
of $S$ become dual after twisting the gluing by $\iota^a$.
Denote the resulting threefold by
$X_1$.
The strict transform of $S$ in $X_1$ is naturally isomorphic to $S$.
The two threefolds $X_1$ and $X_2$ are glued along their copies of $S$
by the automorphism $\iota^a$. This gives a normal crossing
threefold
$$
X_0(a):=X_1\cup_{\iota^a}X_2.
$$
The blow-up centers are chosen so that
$$
N_{S/X_1}\otimes (\iota^a)^*N_{S/X_2}\simeq\mathcal O_S.
$$
Thus $X_0(a)$ is $d$-semistable and so it is smoothable to
a smooth Calabi--Yau threefold
$
X(a).
$

The resulting threefold satisfies
$$
b_2(X(a))=a+3
$$
and
$$
e(X(a))=-256a^2+32a-224.
$$
Hence the second Betti numbers of these Calabi--Yau threefolds are
unbounded as $a$ increases. The authors also show that $X(a)$ is simply connected and
non-K\"ahler.

In \cite{Lee11}, Calabi--Yau threefolds with
$
h^{1,1}=h^{1,2}=1
$  were constructed by the smoothing method. 
Their Hodge diamond is
$$
\begin{array}{ccccccc}
&&&1&&&\\
&&0&&0&&\\
&0&&1&&0&\\
1&&1&&1&&1\\
&0&&1&&0&\\
&&0&&0&&\\
&&&1&&&
\end{array}.
$$

The construction starts with a smooth quartic K3 surface
$
D\subset \mathbb P^3.
$
Choose a sequence
$$
\Gamma=(\gamma_1,\ldots,\gamma_{20})
$$
of smooth rational curves on $D$ such that their classes are linearly
independent in $\operatorname{Pic}(D)_{\mathbb Q}$ and
$$
\gamma_1+\cdots+\gamma_{20}\sim 8H|_D,
$$
where $H$ is the hyperplane class of $\mathbb P^3$.
Starting from $\mathbb P^3$, blow it up successively along the curves
$\gamma_1,\ldots,\gamma_{20}$. Denote the resulting threefold by
$X_\Gamma$. Since all the curves lie on $D$, the strict transform of
$D$ is still isomorphic to $D$ and is an anticanonical divisor in
$X_\Gamma$.

Take another copy
$
Y:=\mathbb P^3
$
and glue $X_\Gamma$ and $Y$ along their copies of $D$. This gives a
normal crossing threefold
$$
Z_\Gamma:=X_\Gamma\cup_D Y.
$$
The condition
$$
\gamma_1+\cdots+\gamma_{20}\sim 8H|_D
$$
implies
$$
N_{D/X_\Gamma}\otimes N_{D/Y}\simeq\mathcal O_D.
$$
Thus $Z_\Gamma$ is $d$-semistable and $Z_\Gamma$ can be smoothed to a simply connected
Calabi--Yau threefold
$
M_\Gamma.
$

For a general sequence of $r$ curves, the Hodge numbers of the smoothing
are given by
$$
h^{1,1}(M_\Gamma)
=
r-\operatorname{rk}\langle\gamma_1,\ldots,\gamma_r\rangle+1
$$
and
$$
h^{1,2}(M_\Gamma)
=
21+\sum_{i=1}^r g(\gamma_i)
-\operatorname{rk}\langle\gamma_1,\ldots,\gamma_r\rangle.
$$
In \cite{Lee11}, a configuration of twenty smooth rational curves was
constructed whose classes are linearly independent and whose sum is
linearly equivalent to $8H|_D$. Since $r=20$, all the curves are rational, and their classes have
rank $20$, the above formulas give
$$
h^{1,1}(M_\Gamma)=20-20+1=1
$$
and
$$
h^{1,2}(M_\Gamma)=21-20=1.
$$
A concrete configuration is obtained on the Fermat quartic surface
$$
D=\{x^4+y^4+z^4+w^4=0\}\subset\mathbb P^3.
$$
Among the lines and conics on this surface, the author chooses eight
lines and twelve conics whose sum is linearly equivalent to $8H|_D$
and whose classes are linearly independent.
If $\xi$ is a generator of $H^2(M_\Gamma,\mathbb Z)$ with
$\xi^3>0$, then
$$
\xi^3=2,\qquad \xi\cdot c_2(M_\Gamma)=44.
$$
The author showed that the threefold $M_\Gamma$ is  simply-connected, Moishezon and that it carries a big line bundle
which induces a rational map
$$
M_\Gamma\dashrightarrow\mathbb P^3
$$
of degree two.

\section{Construction of Calabi--Yau manifolds of higher dimensions -- first non-K\"ahler examples}

Although many non-K\"ahler Calabi--Yau threefolds were known, no
non-K\"ahler Calabi--Yau manifold of dimension greater than three
had previously been constructed.
The smoothing method was subsequently used to construct such
higher-dimensional examples.

In \cite{Lee1}, the first non-K\"ahler Calabi--Yau fourfold was
constructed by smoothing a  normal crossing variety. The
construction uses Beauville's rigid Calabi--Yau threefold together
with two suitably chosen involutions.

Let $E_\zeta$ be the elliptic curve with complex multiplication by a
primitive cube root of unity $\zeta$. The scalar multiplication by
$\zeta$ acts diagonally on $E_\zeta^3$ and has exactly $27$ fixed
points. The quotient
$
E_\zeta^3/\langle\zeta\rangle
$
therefore has $27$ isolated quotient singularities. Blowing up these
singular points gives Beauville's Calabi--Yau threefold $Y$. It is a
projective rigid Calabi--Yau threefold with
$
h^{1,1}(Y)=36, h^{1,2}(Y)=0.
$
The author then chooses two involutions $\sigma_1$ and $\sigma_2$ of
$E_\zeta^3$, induced by matrices with entries in
$\mathbb Z[\zeta]$. They commute with the scalar action of $\zeta$ and
therefore descend to involutions
$$
\rho_1,\rho_2\colon Y\longrightarrow Y.
$$
Each involution acts by multiplication by $-1$ on the holomorphic
three-form of $Y$, and its fixed locus is a smooth surface. An
important feature of the choice is that the two involutions together
generate an infinite group of automorphisms.

For each $i=1,2$, take the product $Y\times\mathbb P^1$. Choose an
involution $\psi$ of $\mathbb P^1$ with two fixed points, and consider
the quotient
$$
(Y\times\mathbb P^1)/(\rho_i\times\psi).
$$
This quotient is singular along smooth surfaces arising from the
fixed loci of $\rho_i$ and the two fixed points of $\psi$. Blowing up
the singular locus gives a smooth projective fourfold $X_i$.

Choose a point $p\in\mathbb P^1$ which is not fixed by $\psi$. The two
fibers $Y\times\{p\}$ and $Y\times\{\psi(p)\}$ are exchanged by the
involution and hence have a common image in the quotient. Its strict
transform in $X_i$ is a smooth divisor $D_i$ isomorphic to $Y$.
Because $\rho_i$ acts by $-1$ on the holomorphic three-form, the
quotient construction makes $D_i$ an anticanonical divisor of $X_i$.
Moreover, its normal bundle in $X_i$ is trivial.

The two fourfolds are now glued transversally along the identifications
$$
D_1\simeq Y\simeq D_2.
$$
This gives a  normal crossing fourfold
$$
X:=X_1\cup_Y X_2.
$$
Since the common divisor is anticanonical in both components, the
dualizing sheaf of $X$ is trivial. The normal bundles on the two sides
are both trivial, so the $d$-semistability condition is automatically
satisfied. So we have
a smoothing of $X$ to a smooth compact complex fourfold $M$. The
triviality of the canonical bundle and upper semicontinuity imply
$$
K_M\simeq\mathcal O_M,\qquad
H^i(M,\mathcal O_M)=0\quad (1\leq i\leq3).
$$
The topology of the smoothing is obtained by gluing
$X_1\setminus D_1$ and $X_2\setminus D_2$ along punctured
neighborhoods of the common divisor. Since both complements are
simply connected, the Seifert--van Kampen theorem shows that $M$ is
simply connected. Hence $M$ is a Calabi--Yau fourfold.

It remains to prove that $M$ is non-K\"ahler. Suppose that $M$ were
K\"ahler. Since a K\"ahler Calabi--Yau manifold of dimension greater
than two is projective, the degeneration would produce a big line
bundle on the normal crossing variety $X$. Its restrictions to the
two components would induce a big divisor class on the common
Calabi--Yau threefold $Y$ which is invariant under both $\rho_1$ and
$\rho_2$.
Pulling this class back to the abelian threefold $E_\zeta^3$ would
give a big, and therefore ample, divisor class invariant under both
$\sigma_1$ and $\sigma_2$. However, the subgroup generated by these
two involutions is infinite, and the intersection of their invariant
Néron--Severi subgroups contains no ample class. This contradiction
shows that $M$ cannot be K\"ahler.

Thus the construction gives a simply connected non-K\"ahler
Calabi--Yau fourfold. Its topological Euler number is
$$
e(M)=108.
$$

Almost concurrently, Sano constructed non-K\"ahler Calabi--Yau manifolds
with arbitrarily large second Betti number in every dimension
$N\geq4$  by a smoothing method
\cite{Sa1}.
This construction is a higher-dimensional analogue of the
Hashimoto--Sano construction discussed in the previous section
\cite{HaSa}.

Let $n\geq 2$, and let
$$
S\subset \mathbb P^2\times\mathbb P^1
$$
be a general hypersurface of bidegree $(3,1)$. Then $S$ is a rational
elliptic surface: the projection to $\mathbb P^1$ gives an elliptic
fibration, while the projection to $\mathbb P^2$ identifies $S$ with
the blow-up of $\mathbb P^2$ at the nine base points of a general
pencil of plane cubic curves.

By repeatedly applying quadratic Cremona transformations to
$\mathbb P^2$, the author constructs, for every positive integer $m$,
another rational elliptic surface $S_m$ and an isomorphism
$
\phi_m\colon S\longrightarrow S_m
$
over $\mathbb P^1$. Although $S$ and $S_m$ have the same elliptic
fibration, the isomorphism $\phi_m$ acts nontrivially on their divisor
classes. The complexity of this action grows with $m$, and this is the
source of the unbounded second Betti numbers.

Next, let
$
T\subset\mathbb P^1\times\mathbb P^n
$
be a general hypersurface of bidegree $(1,n+1)$. Consider
$
Y_1:=\mathbb P^2\times T.
$
Inside $Y_1$, the intersection with $S\times\mathbb P^n$ is naturally
isomorphic to the fiber product
$
D_1\simeq S\times_{\mathbb P^1}T.
$
This is a Calabi--Yau manifold of dimension $n+1$ of Schoen type. In
the same way, a second copy $Y_2=\mathbb P^2\times T$ contains
$$
D_2\simeq S_m\times_{\mathbb P^1}T.
$$
The isomorphism $\phi_m$ induces an isomorphism
$$
\Phi_m\colon D_1\longrightarrow D_2.
$$

The normal bundles of $D_1$ and $D_2$ do not initially satisfy the
$d$-semistability condition. To correct this, the author blows up
$Y_1$ along $m$ smooth divisors lying over smooth elliptic fibers of
$S\to\mathbb P^1$, and then along one additional smooth subvariety
lying over a suitably chosen curve on $S$. Denote the resulting
variety by $X_1$, and put
$
X_2:=Y_2.
$
The blow-up centers are chosen precisely to cancel the product of the
two normal bundles after gluing.

The two components are then glued along the strict transform of $D_1$
and $D_2$ by the isomorphism induced by $\Phi_m$. This gives a simple
normal crossing variety
$$
X_0(m):=X_1\cup_{\Phi_m}X_2.
$$
The double locus is anticanonical in both components, and the
construction ensures that $X_0(m)$ is $d$-semistable and has trivial
dualizing sheaf. Therefore, the Kawamata--Namikawa smoothing theorem
gives a smoothing of $X_0(m)$ to a smooth Calabi--Yau manifold
$
X(m)
$
of dimension $N=n+2$.

The resulting manifold is simply connected and non-K\"ahler. Its
second Betti number is
$$
b_2(X(m))=
\begin{cases}
m+10,& N=4,\\
m+2,& N\geq 5.
\end{cases}
$$
Thus, for every fixed dimension $N\geq4$, the construction produces
Calabi--Yau manifolds with arbitrarily large second Betti number.

The projection to $T$ extends through the smoothing and gives a K3
fibration
$$
X(m)\longrightarrow T.
$$
For a very general smoothing, every meromorphic function on $X(m)$
essentially comes from the base $T$. Consequently, its algebraic
dimension is
$
a(X(m))=\dim T=N-2.
$
Hence these examples are non-K\"ahler Calabi--Yau manifolds whose
algebraic dimensions have codimension two.

\section{Construction of mirror pairs of Calabi--Yau threefolds}

Mirror symmetry is another central geometric phenomenon in the study
of Calabi--Yau manifolds. In the toric setting, mirror Calabi--Yau
threefolds can be constructed from dual reflexive polytopes. For
Calabi--Yau threefolds obtained by smoothing normal crossing
varieties, however, it had remained unclear how a corresponding
mirror partner should be constructed.

A construction scheme for mirror pairs of such Calabi--Yau
threefolds was introduced in \cite{Lee20}. 
A quasi-Fano threefold $X$ is a smooth projective threefold whose
anticanonical linear system contains a smooth K3 surface $D_X$ and
whose higher structure-sheaf cohomology vanishes. If the normal bundle
of $D_X$ in $X$ is trivial, the anticanonical system defines a K3
fibration over $\mathbb P^1$. Two copies of $X$ can then be glued
along $D_X$ to form
$$
X\cup_{D_X}X.
$$
When the $d$-semistability condition is satisfied, this normal
crossing variety can be smoothed to a Calabi--Yau threefold, denoted
by $\Xi_X$. Its Hodge numbers are determined by the cohomology of $X$
and by the image of
$\operatorname{Pic}(X)$ in $\operatorname{Pic}(D_X)$.

The guiding principle comes from Landau--Ginzburg mirror symmetry.
Removing an anticanonical fiber from $X$ produces a noncompact variety
equipped with a natural superpotential. This suggests regarding two
quasi-Fano threefolds $X$ and $Y$ as mirrors when each is related to
the Landau--Ginzburg model of the other and when the lattices induced
on their anticanonical K3 surfaces form a mirror pair of
lattice-polarized K3 surfaces. A basic numerical consequence is
$$
\alpha_X+\alpha_Y=20,
$$
where $\alpha_X$ is the rank of the Picard lattice on $D_X$ induced
from $X$. For such a quasi-Fano mirror pair, the corresponding
smoothings satisfy
$$
h^{1,1}(\Xi_X)=h^{1,2}(\Xi_Y),
\qquad
h^{1,2}(\Xi_X)=h^{1,1}(\Xi_Y).
$$

The main explicit construction begins with a three-dimensional
reflexive polytope $\Delta$. It determines a Gorenstein toric Fano
threefold and a smooth crepant resolution $X(\Sigma_\Delta)$ carrying
a smooth anticanonical K3 surface. Blowing up a suitable smooth curve
on this K3 surface produces a quasi-Fano threefold $X_\Delta$ whose
K3 fiber has trivial normal bundle. The smoothing of
$$
X_\Delta\cup_D X_\Delta
$$
then gives a Calabi--Yau threefold $\Xi_{X_\Delta}$.

It is tempting to apply the same construction directly to the polar
dual polytope $\Delta^\circ$, but this does not in general produce the
mirror partner. The difficulty is that some divisor classes on the
anticanonical K3 surface do not come from the ambient toric
threefold. These missing classes can be recovered by constructing a
different quasi-Fano threefold $Y_{\Delta^\circ}$ through successive
blow-ups along certain irreducible curves on the K3 surface. This
modification is motivated by a fundamental result on mirror symmetry
for K3 surfaces in Gorenstein toric threefolds (\cite{Fa}). The resulting
Picard lattice is the K3-mirror lattice of the lattice associated with
$X_\Delta$. Consequently, the smoothings
$\Xi_{X_\Delta}$ and $\Xi_{Y_{\Delta^\circ}}$ exhibit the expected
exchange of Hodge numbers.

The same idea extends to normal crossing varieties with different
components,
$$
\mathcal X=X_1\cup_D X_2,
\qquad
\mathcal Y=Y_1\cup_{D^\circ}Y_2.
$$
If each pair $(X_i,Y_i)$ is a quasi-Fano mirror pair and the induced
Picard lattices satisfy suitable compatibility conditions, then the
smoothings of $\mathcal X$ and $\mathcal Y$ again satisfy the
Calabi--Yau Hodge-number mirror relation.

Applying this construction to all three-dimensional reflexive
polytopes produces $6518$ mirror pairs of Calabi--Yau threefolds,
including $79$ self-mirror examples \cite{LeeD}. The resulting
families also include the Borcea--Voisin mirror pairs associated with
non-symplectic involutions on K3 surfaces.

The same viewpoint suggests possible higher-dimensional analogues.
In higher dimensions, the expected mirror relation is formulated
primarily in terms of Euler characteristics. New difficulties arise,
however, because smooth crepant resolutions and smooth anticanonical
sections need not exist. A satisfactory extension would therefore
require a broader theory involving singular Calabi--Yau varieties and
their smoothings.

These pairs remain conjectural as mirror pairs in the full sense of
mirror symmetry, since the established result is the exchange of
Hodge numbers. Nevertheless, the construction provides a concrete
framework in which smoothing theory, Landau--Ginzburg models, toric
geometry, and mirror symmetry for lattice-polarized K3 surfaces
interact naturally.

\section{What's  next?}

It is a long-standing open problem whether K\"ahler Calabi--Yau threefolds occur in only finitely
many topological types. In dimension two, the Calabi--Yau manifolds are $K3$ surfaces and
the topological type is unique, but in dimension three no such finiteness result is known.
Large families of K\"ahler Calabi--Yau threefolds are known from toric constructions (\cite{MaHa}), yet
they are believed to realize only finitely many topological types and it is known that
 elliptically fibered K\"ahler
Calabi--Yau threefolds form only finitely many deformation families
(\cite{FiHaSv}).

The examples discussed above show that the smoothing method is
sufficiently flexible to produce non-K\"ahler Calabi--Yau threefolds
with unbounded $b_2$. The unresolved issue is whether this flexibility
of the smoothing method can produce infinitely many topological types of K\"ahler Calabi--Yau threefolds.
The main challenge is to maintain both projectivity and
$d$-semistability of the normal crossing varieties simultaneously.


\end{document}